# PEMANFAATAN TEKNOLOGI SISTEM INFORMASI GEOGRAFIS UNTUK MENUNJANG PEMBANGUNAN DAERAH

Fedro Antonius Pardede, Spits Warnars H.L.H
Fakultas Teknologi Informasi, Universitas Budi Luhur
Jl. Petukangan Utara, Kebayoran Lama, Jakarta Selatan 12260, Indonesia
ch_4nw4r@yahoo.com, spits@bl.ac.id

Abstract :
The territory development will depend on that territory itself, where the word of autonomy for each province or territory will give contribution how Indonesian will responsible for development their territory. In order to develop territory, the information technology can be used as a boost or tools to give and deliver the best information and Geographic Information System is one of the information technology tools which can be used to push every each territory to speed the territory development. As a tool Geographic Information System has an ability to save, process, analysis and deliver information right in time and help the decision maker to make better decision.

## 1. Pendahuluan

Dalam sepulah tahun ke depan, secara lambat tetapi pasti pengembangan SIG akan bergeser dari kegiatan yang bersifat pasif, pengumpulan data digital menjadi kegiatan aktif dinamis berupa penganalisaan data geografis. Untuk itu, data geografis yang dikelola oleh suatu instansi harus dapat diakses dengan mudah oleh instansi lainnya atau pun masyarakat, sehingga keberadaannya akan semakin optimal. Berbagi pakai data (data sharing) merupakan suasana kondusif untuk terciptanya suatu sistem yang interoperability. Suasana keterbukaan ini sangat menunjang keberhasilan implementasi SIG di Indonesia. Beberapa manfaat positif dari penggunaan teknologi SIG seperti efisiensi dan efektifitas, dapat dimanfaatkan untuk kepentingan pembangunan daerah, demi sebesar-besar nya kemakmuran rakyat.

Sistem Informasi Geografis (Geografic Information System - GIS) sebagai tool untuk menyimpan/mengelola, mengolah/menganalisis, dan menyajikan informasi mulai berkembang sejak akhir tahun 1980-an. Untuk penggunaan dan aplikasi SIG di masa depan tiga komponen di atas secara umum masih tetap mendominasi kegiatan utama SIG. Perubahan akan terjadi hanya dalam hal yang terkait dengan pergeseran kepentingan dan implementasi/pemanfaatannya dari ketiga komponen SIG di atas [Briggs, 1999].

Pada awal perkembangannya teknologi SIG ini ditekankan pada pengumpulan dan konversi data dari sistem peta cetak (hardcopy) dan data tabular/numerik (data statistik, dll.) yang terkait ke suatu sistem basis data spasial digital (softcopy). Untuk masa yang akan datang, terutama di negara-negara maju penekanan diharapkan lebih kepada analisis data. Hal ini sangat logis, jika data yang dibutuhkan sebagai basis sudah tersedia dengan baik dan memadai maka pemanfaatan SIG selanjutnya harus lebih ditekankan kepada analisis data untuk memperoleh informasi yang lebih variatif. Walaupun demikian, pekerjaan pengumpulan data tetap harus dilakukan secara terus-menerus, dengan kapasitas yang lebih kecil, untuk tujuan pendinian (updating) data yang sudah ada (Lihat Gambar 1). Penekanan akan lebih diutamakan juga ke arah analisis yang dinamis dan aktif seperti pemodelan dan visualisasi dari data yang dipunyai.

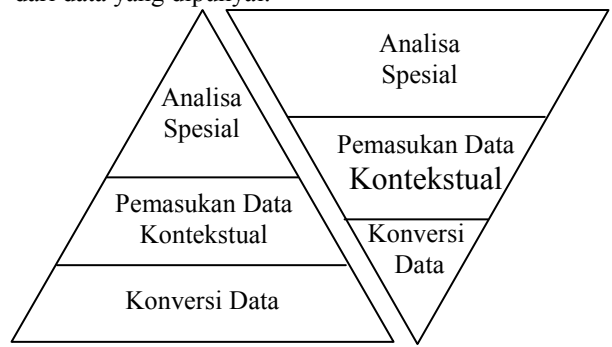

Gambar 1
Perubahan Pemanfaatan SIG

Dari segi ekonomis, proses perkembangan ini akan turut mempengaruhi biaya dan investasi yang harus dikeluarkan dan ditanamkan untuk pembentukan dan pemanfaatan SIG pada tahap selanjutnya. Adapun untuk tahapan berikutnya biaya yang diperlukan akan berkurang pada saat SIG sudah terbentuk dan lebih ditekankan untuk memperoleh informasi sesuai data yang telah tersedia. Demikian juga halnya dengan sumber daya manusia yang diperlukan untuk menangani SIG ini akan semakin kecil karena hanya dibutuhkan sejumlah tenaga untuk memperbaharui data saja dan tenaga ahli untuk menganalisis data sesuai informasi yang diperoleh. Sekedar gambaran, Secara umum apabila pemanfaatan SIG ini dibandingkan dengan sistem pemanfaatan teknologi konvensional maka akan terlihat seperti pada Gambar 2. Pada sistem konvensional makin lama penerapan dilakukan, maka biaya yang harus dikeluarkan akan makin mahal. Hal sebaliknya akan terjadi pada penerapan dengan menggunakan teknologi SIG, biaya yang diperlukan makin lama akan makin murah.

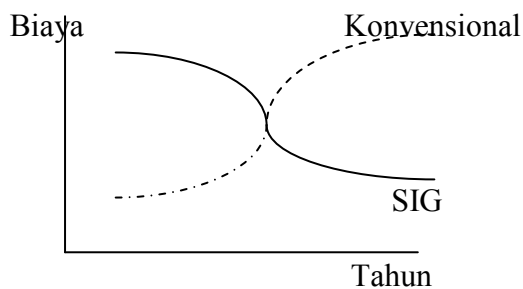

Gambar 2
Perbandingan Antara Penerapan Sistem Konvensional
Dan SIG

## 2. Manfaat Teknologi SIG

Ada dua faktor utama yang terkait dengan masalah keberhasilan implementasi SIG. Kedua hal tersebut yaitu masalah teknologi dan masalah kondisi pengoperasian SIG itu sendiri. Keduanya berhubungan erat dan tidak dapat dipisahkan satu sama lain.

Keberhasilan dari implementasi teknologi SIG sehingga sesuai seperti yang diharapkan akan memberikan dampak yang positif dalam sistem pengelolaan informasi yang menyangkut antara lain masalah efisiensi dan efektifitas, komunikasi yang tepat dan terarah, serta data sebagai aset yang berharga [Briggs, 1999]. Efisiensi dan Efektifitas sistem kerja sebagai dampak dari keberhasilan implementasi teknologi SIG akan semakin terasa. Pada era globalisasi, setiap institusi pada sektor swasta (private sector) dapat bergerak dengan efektif dan efisien setelah mereka menerapkan teknologi SIG untuk membantu pekerjaan mereka di berbagai sektor, bidang atau industri jasa yang mereka tekuni. Kunci kesuksesan bisnis pada sektor ini di masa depan, terutama dalam menghadapi persaingan bebas, adalah adanya sistem pengelolaan yang efisien dan sistem pelayanan yang baik untuk para pelanggan[Longley, 2005]. Sebagai contoh, di suatu negara maju orang memanfaatkan SIG untuk menentukan jalur (route) yang singkat/optimum untuk pengantaran barang dari pabrik ke tempat distributor. Jalur yang singkat tentunya akan menghemat waktu dan biaya pengiriman, sehingga hal ini akan meningkatkan efisiensi dan menjadi pekerjaan mereka menjadi lebih efektif. Di sektor pemerintah (public sector) indikator kesuksesan implementasi SIG akan terletak pada kualitas pelayanan pada masyarakat [Awalin, 2003] atau komunikasi dengan pengguna. Komunikasi ini mungkin lebih kepada pelayanan dalam memberikan informasi yang dibutuhkan masyarakat secara mudah dan cepat. Contohnya menunjukkan arah perjalanan, informasi kepemilikan tanah, lokasi wisata dan lain sebagainya.

Dengan SIG yang baik maka pelayanan informasi yang sifatnya demikian akan dapat secara mudah dan cepat diberikan. Komunikasi Informasi yang Tepat dan Terarah. Dalam suatu sistem informasi yang ideal, penampilan data yang diperlukan harus disesuaikan dengan tingkatan/level dari pemakai (level of users). Tampilan SIG untuk tingkatan Kepala Daerah Propinsi (gubernur) akan berbeda dengan tingkatan pejabat suatu dinas di level kabupaten karena informasi yang diinginkan sudah tentu berbeda. Pada tingkatan dinas di kabupaten, informasi yang diperlukan akan lebih rinci, misalnya seluruh data hasil musim panen harus dapat diketahui untuk setiap kecamatan, sedangkan untuk seorang gubernur informasi ini cukup untuk setiap kabupaten saja. Walaupun tidak tertutup kemungkinan untuk memberikan informasi yang lebih terperinci bagi tingkatan pengguna yang levelnya lebih atas. Terlihat suatu struktur data yang generik sehingga multiguna. Selain itu, untuk kasus data dan informasi yang selayaknya harus diketahui masyarakat umum, seluruh data yang ada pada SIG dapat dibuat dan disusun dalam bentuk sistem jaringan dan memungkinkan untuk dapat disebarluaskan. Dengan demikian memungkinkan masyarakat umum dapat mengakses sendiri data yang ada dan menyimpan sesuai keperluannya dengan/atau tanpa biaya (tergantung kebijaksanaan). Informasi sebagai Aset Data yang dikumpulkan dan dikelola di dalam SIG ini merupakan suatu bentuk aset tersendiri yang tidak berbeda dengan bangunan, mesin-mesin, dan barang-barang inventaris lainnya yang dimiliki oleh suatu institusi. Dalam situasi yang demikian diperkirakan di masa mendatang institusi pemberi jasa informasi termasuk informasi geografis akan lebih berperan. Peranannya akan melebihi perusahaan yang bergerak di bidang perangkat keras (1980-an) dan perangkat lunak (1990-an). Hal ini sangat memungkinkan karena untuk berbagai pengambilan keputusan dalam banyak permasalahan diperlukan informasi (data) yang sampai dengan saat ini belum seluruhnya tersedia dan dapat diperoleh dengan mudah. Sehingga pada akhirnya suatu saat informasi akan menjadi suatu komoditi yang sangat strategis yang banyak dicari dan diminati orang.

## 3. Kondisi dan Manfaat Operasional

Kondisi dan manfaat implementasi SIG di masa depan antara lain menyangkut hal-hal yang berkaitan dengan bidang bisnis dan pemerintah, teknologi komputerisasi, dan terciptanya suasana yang interoperability. Pada Sektor Bisnis dan Pemerintah Dalam era teknologi informasi, kebutuhan untuk jenis pelayanan (informasi) sifat dan penyajiannya sangat ditentukan oleh kebutuhan para pemakai (users requirement), bukan oleh pemberi/penyedia data, seperti halnya kondisi saat ini, karena setiap pemakai memerlukan jenis pelayanan atau informasi yang berbeda. Selain itu, pelayanan atau informasi yang disediakan untuk kebutuhan yang berbeda harus dapat disediakan dalam waktu yang singkat dan dengan biaya yang relatif murah. Karena ringannya biaya untuk memperoleh suatu informasi dengan cepat dan akurat, para pelanggan atau masyarakat pengguna informasi tidak keberatan mengeluarkan biaya untuk mendapatkan informasi yang diperlukan. Persaingan sehat dalam bidang penyedia jasa informasi akan semakin meningkat. Setiap orang akan berusaha untuk menjadi penyedia jasa informasi geografis. Sektor swasta lambat laun akan mengambil alih tugas dan peran dari institusi pemerintah, dan pemerintah secara perlahan dan pasti akan beralih fungsi dari penggerak dan penguasa teknologi menjadi hanya pemakai teknologi. Perangkat lunak SIG yang standar akan lebih populer dibandingkan yang didesain secara khusus. Teknologi Komputerisasi Kebutuhan akan perangkat komputer untuk pengoperasian teknologi SIG akan lebih meningkat. Hal ini disebabkan karena sifat informasi geografis yang dikelola oleh suatu SIG sangat kaya dengan nuansa, mempunyai volume besar dan tersebar (rich and voluminous) [UCGIS, 1998]. Untuk itu diperlukan sistem perangkat keras yang mampu memberikan kecepatan proses data yang tinggi, baik dalam sistem stand-alone maupun jaringan (network), dan dilengkapi dengan media penyimpanan data yang cukup besar (pemanfaatan teknologi terra-byte?). Selain perangkat keras, kemampuan perangkat lunak, baik sistem operasi komputernya sendiri maupun DBMS yang terkait dengan SIG, dituntut untuk semakin canggih (object-oriented) baik dalam hal pengelolaan maupun penyajian data (system multimedia). Hal ini akan menimbulkan persaingan yang cukup ketat di kalangan perusahaan yang bergerak di bidang komputer dan pembuatan perangkat lunak (software house) [Scholten, 1990]. Struktur Informasi Bentuk arsitektur dari jaringan yang tergabung dalam SIG akan memisahkan komputer sebagai pusat basis data dengan komputer sebagai terminal pengolah data. Sehingga perangkat lunak akan mengarah ke sistem modular. Pusat data SIG akan berbagi pakai data (data sharing) dengan pusat data lainnya. Untuk dapat melakukan operasi berbagi pakai data maka dibuat sistem client and server yang terpisah. Setiap pusat data SIG akan bertindak sebagai client. Agar dapat mengakses data dari pusat data SIG lainnya, client-client ini diatur oleh suatu sistem server. Interoperability Hal yang perlu diperhatikan dalam perkembangan SIG ini ialah kemampuan interoperability data. Masalah ini berkenaan dengan sistem penyimpanan data yang digunakan baik data parsial maupun data tekstual. Setiap perangkat lunak SIG memiliki format penyimpanan data grafis dan tekstual tersendiri yang adakalanya tidak dapat dipindahkan ke dalam format lainnya. Dengan perkataan lain data yang ada pada satu SIG tidak dapat digunakan oleh SIG lainnya karena memiliki perbedaan struktur dan format penyimpanan tersendiri. Di masa depan, format data ini (mudah-mudahan) tidak menjadi masalah sepanjang suatu format umum (interface/protocol) telah disetujui sebagai perantara untuk dapat mengubah format yang satu ke format lainnya. Atau paling tidak, akan lebih memudahkan apabila meta-data yang melengkapi data yang ada pada suatu SIG tersedia dan dapat diakses dengan baik. Bakosurtanal mengkoordinasikan SIGNAS (Sistem Informasi Geografis Nasional) yang bertujuan menyusun platform untuk pertukaran data secara nasional. Sebagai catatan, apabila sistem interoperability sudah dapat dicapai maka berbagi pakai data (data sharing) dapat dilaksanakan dengan baik dan akan menguntungkan semua pihak.

## 4. Manfaat SIG Untuk Daerah

Pembangunan daerah di masa depan pada akhirnya akan bergantung kepada daerah itu sendiri. Hal ini disebabkan adanya penerapan otonomi pemerintahan daerah dimana setiap daerah bertanggung jawab untuk dapat mengembangkan daerahnya sesuai dengan potensi dan rencana yang dipunyai. Sejalan dengan itu, sikap para pengambil keputusan pun pada saat ini dituntut untuk lebih terbuka (transparan) sehingga masyarakat dapat mengetahui keputusan dan latar belakang dari kebijakan yang ditetapkan.

Dalam pelaksanaan otonomi, daerah harus menggali dan mengembangkan, secara optimal, potensi dan sumber daya yang ada pada daerahnya demi sebesar-besarnya kemakmuran daerah tersebut. Langkah awal yang harus dilakukan adalah dengan menginventarisasi keberadaan segala sumber daya yang tersedia. Salah satu caranya ialah dengan membangun suatu pusat basis data sumber daya alam dalam media komputer yang terintegrasi dengan SIG. SIG harus tersusun dengan baik dimana semua data daerah, baik data parsial maupun data tekstual, disimpan dan dikelola sehingga untuk memperoleh informasi dapat dilakukan dengan cepat dan tepat. Seperti diketahui, RUTR (Rencana Umum Tata Ruang), baik Kabupaten, Kota maupun Wilayah, merupakan pedoman bagi pemerintah daerah untuk menetapkan lokasi dan manfaat ruang dalam menyusun program-program dan proyek-proyek pembangunan selama jangka waktu tertentu (setahun atau lima tahun). Dalam menyusun RUTR-K/W ini diperlukan data yang menyangkut aspek fisik, sosial dan ekonomi yang berlangsung di daerah tersebut. Dengan diperolehnya data tersebut, potensi/kemampuan, kelemahan, kesempatan dan kendala (Strength, Weakness, Opportunity, Threat) dapat diperkirakan sehingga dapat disusun suatu strategi pengembangan daerah yang efektif dan efisien. Sumber data yang diperlukan diperoleh dari berbagai instansi seperti misalnya Biro Pusat Statistik. Dengan memanfaatkan SIG dimana data yang disimpan tersebut berupa data digital maka informasi yang diperlukan untuk proses perencanaan dapat dilakukan secara mudah dan cepat. Misalkan untuk aplikasi analisis kesesuaian fisik lahan. Salah satu metoda untuk memperoleh inforrnasi kesesuaian lahan ini ialah dengan memberikan score pada setiap jenis data yang digunakan sesuai kondisi data tersebut misalnya jenis tanah, tingkat kemiringan lereng, jumlah curah hujan pertahunnya dan data lain yang ada. Umumnya proses ini diIakukan dengan menggunakan analisis tumpang tindih (overlay) dari seluruh data yang berupa peta-peta tematik sehingga dapat dilakukan penjumlahan score untuk menentukan kesesuaian lahan berdasarkan criteria yang dipakai.

## 5. Pemecahan Masalah Dan Jalan Keluar

Untuk pengembangan SIG daerah dan pengadaan data dasar yang menunjangnya ada beberapa aspek yang harus dikaji dan dapat dicoba dicari jalan keluarnya. Aspek yang pertama menyangkut aspek pendanaan, aspek kedua menyangkut teknologi, dan aspek ketiga menyangkut sumber daya manusia (SDM) yang harus tersedia. Aspek Pendanaan Dari segi pendanaan diperkirakan hampir sebagian besar daerah (propinsi) di Indonesia memiliki sumber daya alam yang dapat digali dan dapat dimanfaatkan sebesar-besarnya dan setertib-tertibnya untuk mendapatkan sumber dana untuk pembangunan dan kesejahteraan daerah. Dalam kaitan dengan otonomi ini pemerintah daerah dapat mencari peluang sumber dana dengan leluasa baik melalui pinjaman luar negeri, penanaman modal dalam negeri (PMDN), penanaman modal asing (PMA), bahkan dana masyarakat untuk menggali dan memanfaatkan sumber daya alam dan sumber daya lainnya untuk sebesar-besar kemakmuran rakyat daerah. Aspek Teknologi Seperti dijelaskan pada uraian-uraian di atas, hal yang sangat mendasar dalam pengembangan SIG adalah ketersediaan data dasar topografis dan tematis yang terkait. Dalam sistem konvensional pengadaan kedua jenis data dasar di atas sangat memakan waktu, tenaga, dan dana. Disadari bahwa selama limapuluh empat tahun merdeka belum seluruh wilayah Indonesia terliput oleh peta dasar yang memadai baik topografis maupun tematis. Salah satu hambatannya adalah masalah teknis untuk menangani pemetaan dengan area yang terpencar-pencar dan begitu luasnya. Dalam sistem pemetaan modern pengadaan kedua jenis data dasar di atas dapat dilakukan dengan mempersingkat waktu pelaksanaan yang cukup signifikan. Dengan memanfaatkan teknologi GPS (Global Positioning System), penyebaran titik kerangka dasar nasional, dan kerangka dasar turunannya di daerah yang merupakan jaringan referensi pengadaan data dasar topografis dapat dilakukan jauh lebih cepat dibandingkan dengan sistem yang ada sebelumnya (sistem terestris dan sistem doppler). Pengambilan data untuk pengadaan data dasar dalam bentuk peta digital, teknologi fotogrametri digital (softcopy photogrammetry) dan sistem CAD dapat mengefisiensikan waktu pemrosesan yang jauh lebih baik daripada sistem sebelumnya (sistem fotogrametri analog dan analitis). Teknologi ini berkembang sejak awal 1990-an dan sampai saat ini sudah ada belasan sistem yang dapat dibeli di pasaran [Leberl, 1991].

Di negara-negara maju sistem fotogrametri digital ini sudah menjadi pilihan utama (hanya satu-satunya pilihan) mengingat sistem sebelumnya (analog dan analitis) sudah tidak dikembangkan dan diproduksi lagi. Untuk daerah-daerah yang sulit dipotret karena cuaca dan liputan awan yang di atas toleransi dapat diatasi dengan cara memanfaatkan satelit radar interferometri yang teknologinya makin lama makin baik, memadai, dan menjanjikan (promising) untuk pengadaan data dasar topografi terutama informasi relief atau data ketinggian. Sehingga masalah pengadaan data dasar untuk daerah-daerah seperti di pedalaman Kalimantan (mudah-mudahan) tidak akan tergantung lagi kepada keadaan cuaca. Untuk keperluan pengadaan data dasar tematis, dapat dilakukan dengan teknologi penginderaan jauh dengan memanfaatkan citra satelit yang resolusinya makin lama makin baik. Satelit inderaja IKONOS [Fritz, 1999] dalam waktu dekat akan diluncurkan dan mempunyai resolusi parsial di bawah satu meter. Dengan demikian detil informasi topografis dan tematis dapat lebih memadai untuk pengadaan data dasar tematis (bahkan mungkin untuk pengadaan data dasar topografis) atau data dasar lainnya yang diperlukan untuk keperluan pembangunan dan pengembangan SIG. Proyek LREP II (Land Resource Evaluation and Planning II) yang diselenggarakan pada tahun 1992-1997 yang dikoordinir oleh Bakosurtanal dengan bantuan Dana Bank Pembangunan Asia (ADB) adalah salah satu contoh proyek yang mencoba menerapkan teknologi modern yang ada kaitannya dengan penataan ruang daerah pada level kabupaten/kota. Proyek ini dapat dikatakan merupakan suatu upaya untuk memberdayakan daerah dalam hal penerapan teknologi pemetaan digital dan SIG dengan melibatkan sumber daya manusia dari daerah yang bersangkutan. Pada setiap daerah di 18 Propinsi dibuatkan proyek percontohan yang berbeda-beda yang dipilih dan ditentukan sendiri oleh masing-masing Bappeda [Hakim, 1996]. Aspek Sumber Daya Manusia Berbicara tentang sumber daya manusia (SDM) daerah dan kaitannya dengan teknologi modern bisa jadi berarti kendala, kesulitan, dan hambatan. Demikian juga SDM untuk SIG akan merupakan barang langka di daerah. Selama ini jumlah tenaga ahli dan tenaga terdidik yang memahami pemanfaatan SIG, dan profesi-profesi lainnya belum mencapai 1.000 orang (BPPT, 1994) dan sebagian besar berdomisili di sekitar Jakarta. Pada saat otonomi daerah dijalankan, apakah tenaga-tenaga ini dapat didistribusi/terdistribusi ke daerah, merupakan suatu pertanyaan besar. Berdasarkan pengalaman LREP II, dalam hal pengadaan SDM khusus yang menangani pemetaan digital dan SIG dapat ditempuh tiga cara, yaitu melalui suatu pendidikan dengan memberikan beasiswa tugas belajar di dalam negeri atau di luar negeri, memberikan pelatihan singkat di dalam negeri dan/atau mengadakan suatu on the job training (OJT) di masing-masing daerah yang memiliki proyek percontohan. Untuk jangka pendek pengadaan SDM dengan cara kedua terakhir dengan bimbingan tenaga ahli yang berpengalaman boleh dikatakan cukup efektif. Pengadaan SDM seperti di jelaskan di atas khusus untuk pemetaan digital dan SIG dapat ditempuh oleh daerah-derah otonom dengan cara meminta bantuan atau mengadakan kerjasama dengan perguruan-perguruan tinggi yang memang memiliki tenaga ahli di bidang ini, antara lain ITB (Jurusan Teknik Geodesi), Universitas Indonesia (Fakultas Geografi), dan Universitas Gajah Mada (Fakultas Geografi).

## 6. Kesimpulan

Sistem Informasi Geografis merupakan suatu sistem informasi yang sangat berguna untuk membantu pengambilan keputusan karena mampu untuk mengelola dan menganalisis data parsial dan tekstual. Dengan demikian, informasi yang dihasilkan tidak hanya informasi tekstual atau deskriptif saja tetapi dapat juga diketahui informasi lokasinya. Teknologi SIG harus sudah dimasyarakatkan terutama kepada setiap daerah. Penggunaan teknologi ini akan lebih menghemat biaya perencanaan pembangunan dibandingkan dengan teknologi konvensional yang masih dipakai saat ini.

Tingkat efisiensi dan efektifitas pelaksanaan pembangunan akan meningkat apabila SIG diaplikasikan untuk perencanaan di segala sektor pembangunan. Selain itu, kualitas pelayanan pada masyarakat dari instansi pemerintah dan swasta akan bertambah baik. Masyarakat akan lebih berperan dalam menentukan jenis informasi yang dibutuhkan dan informasi topografis yang akurat dan terpercaya dapat diperoleh dalam waktu yang singkat.

Peran pemerintah sebagai penyedia jasa informasi pada saat ini akan jauh berkurang dan digantikan oleh sektor swasta. Pemerintah akan menjadi pemakai teknologi karena penggerak dan penguasa teknologi akan diperankan oleh sektor swasta. Komputer yang berkualitas dan berkuantitas dan terhubung satu dengan lainnya (networking) sudah menjadi keharusan untuk mengaplikasikan teknologi SIG dengan sempurna.

Aspek analisis parsial dari teknologi SIG akan lebih berperan mengingat aspek kegiatan pembentukan basis data topografis digital sudah tidak menjadi masalah besar lagi, terutama di negara yang telah maju. Telah banyak instansi, baik pemerintah maupun swasta, yang memanfaatkan keunggulan SIG (Sistem Informasi Geografis) untuk diterapkan pada berbagai bidang studi sehingga membantu dalam pengambilan keputusan. Salah satu cara yang sedang dilakukan ialah dengan membentuk standardisasi SIG (Sistem Informasi Geografis) terutama bentuk penyimpanan data dan sumber data yang digunakan dan dikelola oleh suatu badan (Bakosurtanal).

Sistem Informasi Geografis (SIG) merupakan suatu sistem informasi yang multi-disiplin. Banyak persoalan- persoalan, selain masalah fisik, ekonomi, sosial dan hankam yang memanfaatkannya.Pengembangan Sistem Informasi Geografis (SIG) dapat juga di lihat pada Sistem Operasi Windows Yang Dikenal Dengan Nama Windows Encarta.

**Daftar Pustaka**
Briggs, Ron, (1999), POEC5319 Introduction to GIS, http://www.utdallas/edu/~briggs/poec
 6381/.lecture, BPPT, Bakosurtanal, LAPAN (1994) Direktori Remote Sensing dan SIG di Indonesia, Laporan Tahunan.
Fritz,L.W (1999), Commercial Earth Observation Satellites, CIM International, Vol. 13, No. 5, pp. 6-9. Leberl, Franz and Boulder, (l991). The Promise of Softcopy Photogrammetry, Digital Photogrammetic Systems, Herbert Wichmann Verlag GmbH, Karlsruhe, pp. 3-14.
UCGIS, (1998), Research Priorities for Geographic Information Science, http://www.ucgis.org, 16 pp.
Hakim, D.Muhally, (1996), Laporan Akhir Ahli Basis Data, Proyck LREP II, Bakosurtanal, Cibinong-Bogor.
Sylviawati, V.A., (1994) Aplikasi Sistem Informasi Geografis untuk Model Simulasi Analisis Kebakaran Hutan Tanaman Industri dengan Arc/lnfo", Skripsi, Teknik Geodesi-ITB.
Angkatan 1997 (1998), Studi Pengembangan Wilayah Kabupaten Sukabumi, Jawa Barat", Laporan Studio Perencanaan Wilayah, Magister Planologi-ITB.
Awalin, L.J, Sukojo, B.M. (2003) Pembuatan dan Analisa Sistem Informasi Geografis Distibusi Jaringan Listrik, Makara, Teknologi, Vol 7, No.1, april 2003, pp. 33-44.
Longley, P., Goodchild, M.F., Maguire, D.J., Rhind, D.W. (2005) Geographical Information Systems and Science, John Willey& Sons, 2nd edition.
Scholten, H.J., de Lepper, M.J.C. (1990) The Benefits of the Application of Geographical Information Systems in Public and Environmental Health, Consultation on Epidemiological and Statistical Methods of Rapid Health Assessment, Geneva, 26-29 Nov 1990.

Kingston, R.Carver, S., Evans, A, Turton, I. (2000) web-based public participation geographical information systems : an aid to local environmental decision making, Journal of Computers, Environments and Urban Systems, Vol. 24, Issue 2, pp. 109-125.

Martin,D, Atkinson, P, (2000) Editorial : Innovaton in GIS application ?, Journal of Computers, Environments and Urban Systems, Vol. 24, Issue 2, pp. 61-64.